# Planarized Fabrication Process With Two Layers of SIS Josephson Junctions and Integration of SIS and SFS $\pi$-Junctions

Sergey K. Tolpygo, *Senior Member, IEEE,* Vladimir Bolkhovsky, Ravi Rastogi, Scott Zarr, Alexandra L. Day, Evan Golden, Terence J. Weir, Alex Wynn, and Leonard M. Johnson, *Senior Member, IEEE*

***Abstract*—We present our new fabrication Process for Superconductor Electronics (PSE2) that integrates two (2) layers of Josephson junctions (JJs) in a fully planarized multilayer process on 200-mm wafers. The two junction layers can be, e.g., conventional Superconductor-Insulator-Superconductor (SIS) Nb/Al/AlO$_x$/Nb junctions with the same or different Josephson critical current densities, $J_c$. The process also allows integration of high-$J_c$ Superconductor–Ferromagnet–Superconductor (SFS) or SFS'S JJs on the first junction layer with Nb/Al/AlO$_x$/Nb trilayer junctions on the second JJ layer, or vice versa. In the present node, the SFS trilayer, Nb/Ni/Nb is placed below the standard SIS trilayer and separated by one niobium wiring layer. The main purpose of integrating the SFS and SIS junction layers is to provide compact $\pi$-phase shifters in logic cells of superconductor digital circuits and random access memories, and thereby increase the integration scale and functional density of superconductor electronics. The current node of the two-junction-layer process has six planarized niobium layers, two layers of resistors, and 350-nm minimum feature size. The target critical current densities for the SIS JJs are 100 μA/μm$^2$ and 200 μA/μm$^2$. We present the salient features of the new process, fabrication details, and characterization results on two layers of JJs integrated into one process, both for the conventional and $\pi$-junctions.**

***Index Terms*—Josephson device fabrication, Nb/AlO$_x$/Nb junctions, Nb/Ni/Nb junctions, $\pi$-junctions, SFS junctions, SFQ electronics, superconducting device fabrication, superconducting electronics, superconducting electronics fabrication.**

## I. Introduction

OVER the past six years, our team has developed several nodes of the MIT Lincoln Laboratory (MIT LL) fabrication process for superconductor electronics (SCE), named SFQee, by progressively increasing the number of superconducting layers and decreasing the minimum linewidth [1]-[4]. This process development has been done within the framework of the IARPA C3 Program [5] for applications in energy efficient digital circuits based on different versions of pulse-based Single Flux Quantum (SFQ) logic.

This research is based upon work supported by the Office of the Director of National Intelligence, Intelligence Advanced Research Projects Activity, via Air Force Contract FA872105C0002. *(Corresponding author: Sergey K. Tolpygo.)*

All authors are with the Lincoln Laboratory, Massachusetts Institute of Technology, Lexington, MA 02421, USA (e-mail: sergey.tolpygo@ll.mit.edu).

Our standard SFQ5ee process node has one layer of Nb/Al/AlO$_x$/Nb [6] Josephson junctions (JJs), nine superconducting layers, full planarization on all metal layers, and 350-nm minimum feature size. Using this development approach, we have demonstrated circuits with close to one million JJs and circuit density over 1.3×10$^6$ JJ/cm$^2$ [7],[8]. To further increase the integration scale, we have recently introduced a fabrication process with self-shunted JJs and thin-film kinetic inductors [4],[9] in order to eliminate shunt resistors and reduce the area occupied by circuit inductors, the main impediments to increasing the integration scale of superconductor electronics [10]. The integration scale of SCE can be further increased by implementing multiple layers of active devices – Josephson junctions. A fabrication process with two Nb/Al/AlO$_x$/Nb trilayers with $J_c$ = 100 μA/μm$^2$ was recently demonstrated [11] but applied so far only to fabrication of simple quantum-flux-parametron (QFP) circuits using all JJs of the same, about 1 μm, size.

To grow the integration scale of superconductor electronics and enable high-density, integrated cryogenic memories, it would also be highly desirable to incorporate new materials, in particular magnetic materials and high-*k* dielectrics, into superconductor integrated circuits fabrication. It is well known that JJs with magnetic barriers possess many interesting properties [12] that can be very beneficial for superconductor electronics; see, e.g., [13] for a review. For instance, potential benefits of Josephson π-junctions, i.e., JJs having a π phase change of the superconducting wave function across the barrier, were suggested a long time ago [14]. SFS-type π-junctions [15],[16] can work as phase shifters in logic gates and increase their operation margins, see, e.g., [16]-[21] and references therein. Also, critical current of JJs with magnetic barriers (magnetic JJs) can be varied by changing their internal magnetization. This feature can be used in cryogenic memories, superconducting spintronics, and other applications, see [13], [22], [23] for a review. However, an industrial quality technology for integration of magnetic π-junctions with regular (0-shift) junctions in large-scale integrated circuits has not been developed, despite a considerable amount of research on Josephson junctions with a vast variety of magnetic barriers, and demonstrations of simple circuits.

In this work, we present our new fabrication process PSE2 that integrates two layers of Josephson junctions. These layers can be conventional SIS JJs with the same or different $J_c$ values, or any one of them can be a layer of high-$J_c$ SFS junctions, e.g., Nb/Ni/Nb. In order to characterize the process, we investigated and present below fabrication details and JJ testing results for the following combinations:

a) two layers of Nb/Al/$AlO_x$/Nb JJs with $J_c$ = 100 μA/μm$^2$;
b) self-shunted Nb/Al/$AlO_x$/Nb JJs with $J_c$ = 600 μA/μm$^2$ on the first layer and our standard Nb/Al/$AlO_x$/Nb JJs with $J_c$ = 100 μA/μm$^2$ on the second layer;
c) high-$J_c$ Nb/Ni/Nb (SFS-type) or Nb/Ni/$Mo_2N$/Nb (SFS"S-type) π-junctions on the first layer and Nb/Al/$AlO_x$/Nb JJs with $J_c$ = 100 μA/μm$^2$ on the second.

## II. PSE2 Process Overview

### A. Layer Stack and $J_c$ Options

A cross section of the PSE2 process is shown in Fig. 1a. In this process, interlayer $SiO_2$ dielectric is deposited over each patterned Nb layer and planarized using chemical-mechanical polishing (CMP). This provides smooth, planarized dielectric surfaces for placing JJ trilayers at any convenient level, and enables deep submicron patterning of all layers by 248-nm photolithography and high-density plasma etching.

The process consists of two basic 3-Nb-layer process modules stacked on top of each other. The first module has a Nb ground plane layer (labeled M4 in white in Fig. 1a), a JJ trilayer (J5/M5), a top Nb wiring layer (M6), and a resistor layer (R4). We start numbering Nb layers from M4 in order to reserve numbers for additional wiring layers if they need to be placed below the first ground plane in the future and to preserve the same notations of the lithographic layers as the ones used in our SFQ4ee and SFQ5ee processes.

The second 3-Nb-layer module has the second JJ trilayer (J7/M7), top wiring layer (M8), the second resistor layer (R7), and a Nb ground plane (M9) on top of the stack. The two modules differ by the relative position of the resistor layer and the way resistors are wired. In the 1st module, R4 resistors are below the trilayer and interconnected by the trilayer's base electrode, M5. In the 2nd module, R7 resistors are above the trilayer and interconnected by the top wiring layer, M8 along with the junctions.

The fabrication process of the first module is very similar to our new process SC1 [24], whereas the second JJ module is identical to the SFQee process developed in [1],[2]. Vias to Nb layers M4, M5, etc., are named, respectively, I4, I5, etc. Contact holes to resistors R4 and R7 are named C4 and C7R, and to contacts to junctions J5 and J7 as C5J and C7J, respectively. We also kept, wherever possible, the same thicknesses of metal and dielectric layers in order to preserve the same inductance values. This allowed us to port designs of all test structures, process control monitors (PCMs), and simple circuits from the SFQ5ee process, using a simple remapping.

In the course of process development, we also used a truncated version of the PSE2 process, which lacks one niobium

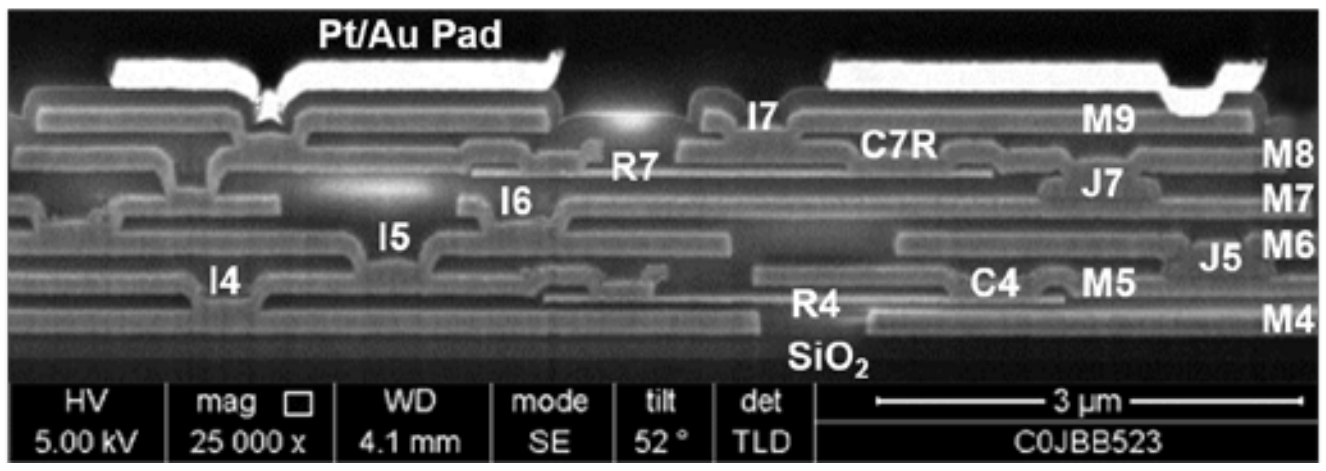

(a)

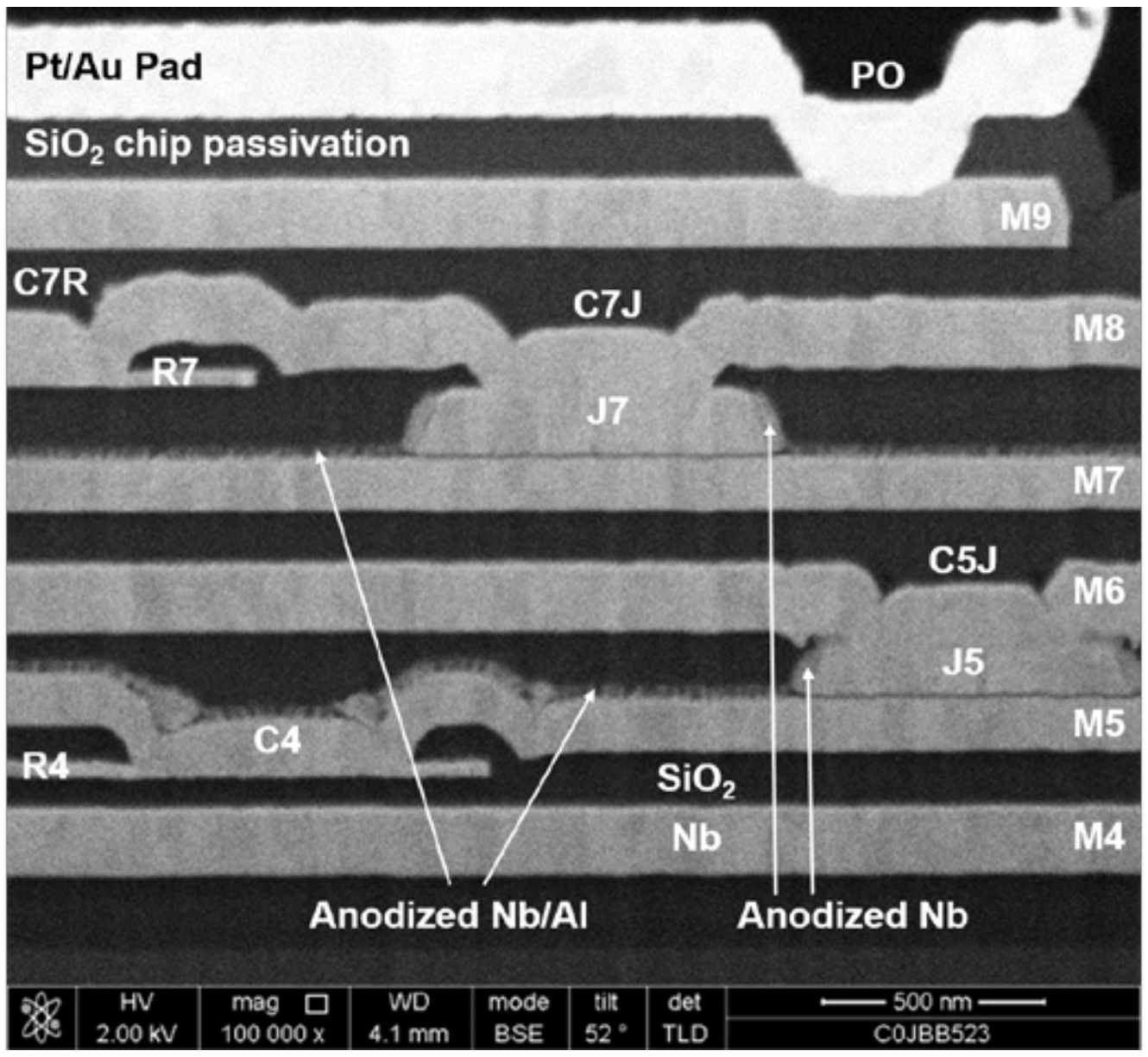

(b)

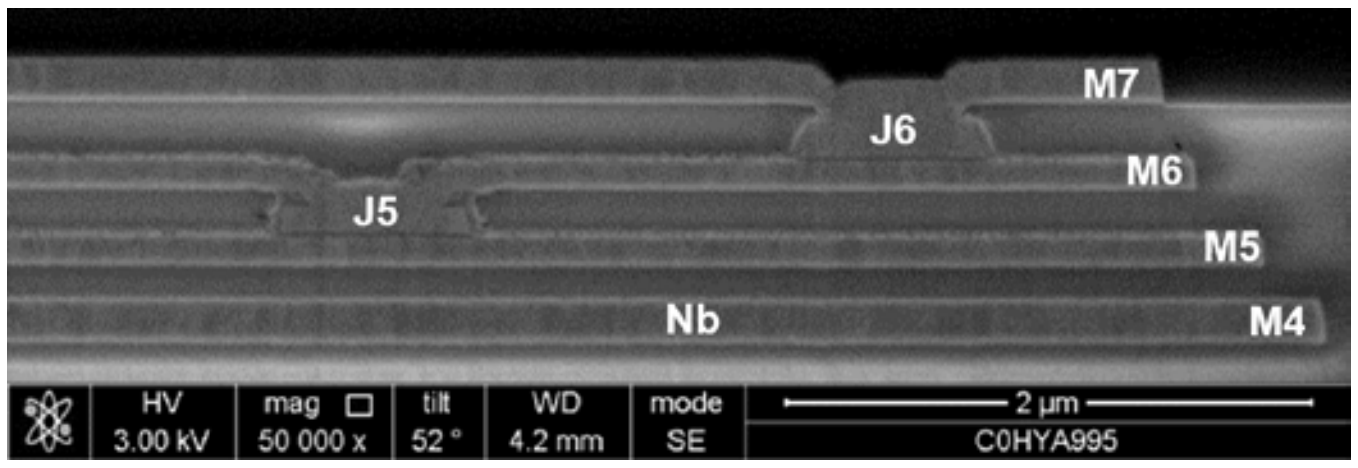

(c)

Fig. 1. (a) A scanning electron microscope (SEM) image of a cross section made by focused ion beam (FIB) of a new process, PSE2 with two layers of Josephson junctions placed near the bottom of the process stack. White labels indicate process layers in notations used in the SFQ5ee process: J5 is the first Josephson junction layer, C5J – top via to junction J5, R4 – the first resistor layer, C4 – via to the resistor R4, I4 – via to Nb layer M4, and so on. M7 is the base electrode of the second trilayer and J7 is the second layer of JJs. Note the difference in placement of the first resistor layer R4, which is below J5, relative to the second resistor layer R7, which is above J7.

(b) A zoom in on the PSE2 cross section, showing 700-nm diameter junctions J5 and J7. The anodic Nb oxide layer is clearly visible around the junctions J5 and J7; also visible on the surface of layers M5 and M7 is a columnar mixed-oxide $(Nb/Al)O_x$ layer formed during anodization of the Al layer on Nb base electrode of the trilayers. The top surface of the wafer is passivated by $SiO_2$ and chip contact pads are metallized by Pt/Au through passivation openings (PO).

(c) Cross section of a structure fabricated by the truncated process PSE2* with two layers of Nb/Al/$AlO_x$/Nb Josephson junctions, J5 and J6, placed above each other. With respect to PSE2, the PSE* process lacks one wiring layer between the two trilayers, labeled M6 in (a) and (b). Resistors and the fifth Nb layer are not visible in this test structure. Junctions J5 and J6 are circular JJs with a diameter of 700 nm. The PSE2* was used in some process development runs in order to investigate possible proximity effects between JJs on the adjacent layers.

layer between the first and the second layer of junctions, as shown in Fig. 1(b). We will refer to this process as PSE2*. This process was used in some of the runs in order to investigate possible proximity effects between JJs on adjacent layers, J5 and J6 in Fig. 1b.

We investigated and suggest for implementation the following combinations of Josephson junction layers: a) two layers of Nb/Al/AlO$_x$/Nb with the same target $J_c$ values of 100 μA/μm$^2$; b) two layers of Nb/Al/AlO$_x$/Nb with different $J_c$ values: 600 μA/μm$^2$ on the 1$^{st}$ layer and 100 μA/μm$^2$ on the 2$^{nd}$ layer; c) Nb/Ni/Nb π-junctions on the 1$^{st}$ junction layer and Nb/Al/AlO$_x$/Nb JJs with $J_c$ = 100 μA/μm$^2$ on the 2$^{nd}$ JJ layer. We prefer to place the higher-$J_c$ trilayer at the bottom of the process stack because its thinner barrier is more sensitive to the surface quality and because it requires smaller-size JJs, which are easier to fabricate on the lower levels.

For the *SFS* trilayer, we have investigated only its placement as the first JJ layer at the bottom of the process stack. Since the intended use of the *SFS* junctions is as passive π-phase shifters and not switching JJs, there are no strict requirements to the exact value and spread of their critical currents, $I_c$. The JJs only must be in the π-junction regime and must have much larger $I_c$ than other JJs in the logic, in which case the Josephson inductance of the π-junctions can be neglected in comparison with the geometrical and kinetic inductances of the cell inductors. These relaxed requirements to $I_c$ simplify fabrication of π-junctions and could allow use of the *SFS* layer as the second JJ layer if this becomes more convenient from the design standpoint.

A summary of the PSE2 salient features is given in Table I.

### *B. SIS and SFS Trilayers Fabrication*

Our fabrication process for Nb/Al/AlO$_x$/Nb junctions was described in detail in [1]–[3],[24]. We adjusted the Nb deposition process in order to obtain a more uniform stress distribution, primarily between the center and edge of the wafers. This was achieved by decreasing the sputtering power below 1 kW and lowering the Ar pressure to about 0.13 Pa [23].

For the fabrication of π-junctions, we used Nb/Ni/Nb trilayers deposited in-situ on 200-mm wafers in an Endura PVD cluster tool (Applied Materials, Inc.) Based on the recent results on Nb/Ni/Nb junctions [25],[26], the π-junction regime spans the range of Ni thicknesses from about 0.8 nm to about 3.4 nm. In our process, we target the center of this range, using trilayers with the nominal thickness of Ni barrier $t_{Ni}$ = 2, 2.5, and 3 nm.

In some experiments we also implemented Nb/Ni/Mo$_2$N/Nb junctions, making a SFS'S-type structure, where a 35-nm overlay of superconducting Mo$_2$N [4] was deposited on the Ni surface prior to the Nb top electrode deposition. This amorphous overlay was used to decrease the Josephson critical current density of magnetic junctions by modifying the interface with the Nb top electrode.

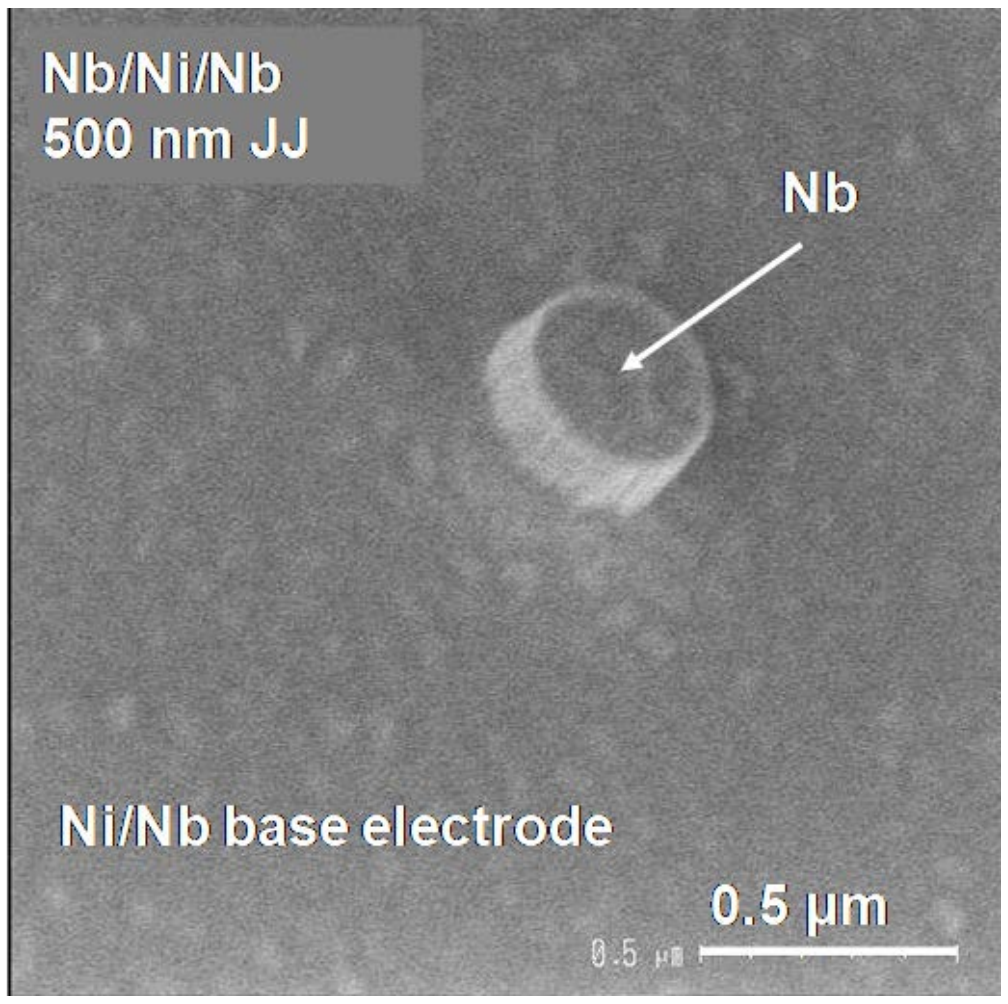

(a)

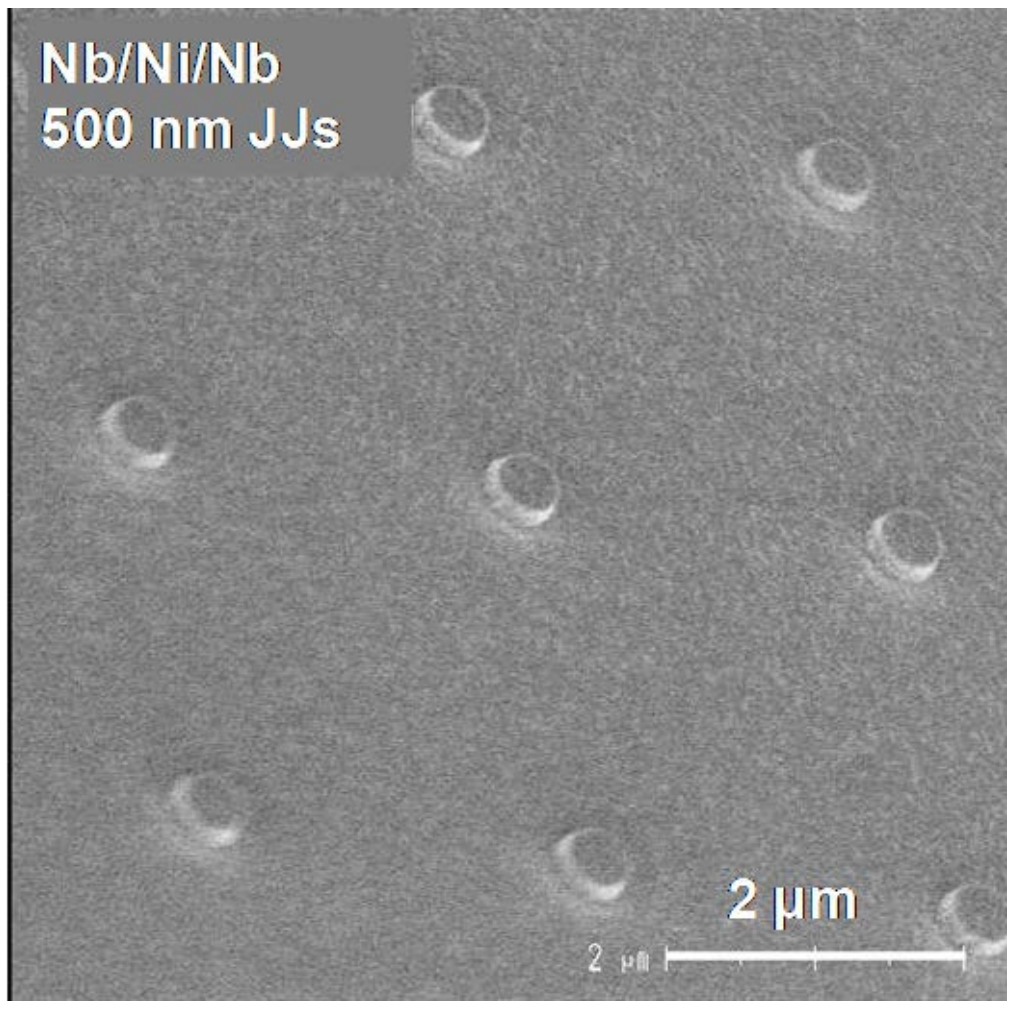

(b)

Fig. 2. SEM images of Nb/Ni/Nb Josephson junctions with 500-nm diameter after photolithography and etching of the top Nb electrode. (a) A single 500-nm Nb/Ni/Nb junction; (b) arrays of 500-nm Nb/Ni/Nb junctions. SEM images were taken at a 45° wafer tilt.

### *C. 248-nm Photolithography*

For 248-nm photolithography of all process layers, we use a Canon FPA-3000 EX4 stepper and a TEL-12 photoresist processing cluster. We use AR3 bottom antireflection coating (BARC) with 62-nm thickness and UV5-0.6 photoresist with 540-nm thickness for photolithography of all metal layers; UV5-0.8 resist with 690-nm thickness is used for dielectric layers. The exposure dose for metal layers is typically in the range from 120 to 170 J/m$^2$ and is optimized for the required feature size and CD control. For dielectric layers, the dose is in the range from 230 to 270 J/m$^2$.

### *D. Etching: Nb, Ni, Resistors, Interlayer SiO$_2$*

Nb etching is done in a SAMCO RIE-202 cluster tool. We use high-density plasma etching at an RF power of 600 W and bias power of 200 W, a $Cl_2$/Ar mixture with 90/10 sccm flow rates, and 0.7 Pa pressure. The etching end-point is detected using an optical emission spectrograph.

The junction Nb top electrode etching is stopped on Al/$AlO_x$ in case of Nb/Al/$AlO_x$/Nb and on Ni for Nb/Ni/Nb junctions. After dry stripping of the photoresist etch mask, the wafers are anodized in ammonium pentaborate solution in ethylene glycol at 20 V in order to convert the Al layer (or Ni barrier layer) remaining on the exposed surface of the trilayer base electrode into anodic oxides, and to passivate the sidewalls of Nb junctions; see Fig. 1b.

The typical SEM images of the etched Nb/Ni/Nb junctions with diameter $d$ = 500 nm are shown in Fig. 2.

Ni barrier etching required for patterning the base electrode of JJs is done in a Centura BE cluster tool, using a $Cl_2$/Ar mixture and high bias power.

Etching of Mo and $MoN_x$ resistors is done in a Centura BE cluster tool, using an RF power of 600 W, bias power of 60 W, a $SF_6$/$O_2$ mixture at 1.33 Pa, and 60/40 sccm flow rates.

$SiO_2$ etching is done in a SAMCO RIE-200 system, using an RF power of 700 W and bias power of 300 W, a $CF_4$/$CHF_3$/Ar mixture at 3 Pa, and 50/50/50 sccm flow rates.

Photoresist stripping is done in $O_2$ plasma in the SAMCO RIE-202 cluster tool. We usually use an additional wet cleaning after the dry stripping to remove etch residues from the wafer surface, using a commercial remover.

### *E. Interlayer Dielectric Deposition and CMP*

Each patterned junction layer is planarized by depositing a thick layer of $SiO_2$ and polishing it down to the process-required interlayer dielectric thickness by chemical-mechanical polishing (CMP). $SiO_2$ deposition is done in a Sequel (Novellus) PECVD system, using a $SiH_4/N_2O/N_2$ gas mixture at 2.4 Torr, 300/11000/1500 sccm flow rates, 1.1 kW RF power, and about a 4.5 nm/s deposition rate. $SiO_2$ PECVD is done at 150 $^oC$ in order to prevent degradation of the Nb and Josephson junctions. Dielectric CMP is done using a Mirra (Applied Materials, Inc.) polisher with five-zone pressure adjustment. We use silica-based polishing slurry W-2000 and DOW-IC1000 polishing pad (Rodel, Inc.). Polishing rate and uniformity are optimized by adjusting the polisher head pressure and platen/head rotation speed.

## III. Electrical Test Results

### *A. Two Layers of Nb/Al/$AlO_x$/Nb Junctions*

The uniformity of $J_c$ on both junction layers and electrical properties of Josephson junctions were characterized firstly at room temperature by using a wafer prober to measure the resistance of individual JJs on all dies on the wafer, about 400 JJs on each die. Room-temperature resistance $R_{300}$ or conductance $G_{300} = 1/R_{300}$ of JJs and their distributions directly characterize the value and distribution of junction $I_c$ as follows from [27]; see also [1] for more details.

We present statistical data for the smallest (700-nm-diameter) JJs allowed in the process because their parameter variations represent the largest possible variation of JJ parameters in the process. Larger junctions have smaller variations within the die and across the wafer. After dicing the wafers, individual JJs and series array of JJs were measured also at 4.2 K to confirm the results, but on a much smaller number of dies.

It can be seen from the wafermaps and the data presented in Fig. 3 that the first and the second trilayer have very similar properties and similar distributions of JJ parameters across 200-mm wafers. The map shows the mean resistance of 110 JJs on each die, within a 5 mm x 5 mm chip, and the second moment of the resistance distribution, which we, for convenience, refer to as a standard deviation, 1σ (normalized to the mean value). This does not imply a true normality of the distribution based on limited statistics. We note that the resistance wafermaps represent a cumulative effect of a global variation of the tunnel barrier transparency ($J_c$) on the wafer scale, characterized by the mean values, and local variations of the junction area and the barrier within the die, which are characterized by the normalized standard deviation.

The data for a different combination of $J_c$s of the trilayers are shown in Fig. 4. In this case we targeted $J_c$ = 600 μA/μm$^2$ on the first trilayer and 100 μA/μm$^2$ on the second trilayer. The corresponding targets for the resistance $R_{300}$ of 700-nm circular JJs are 7±0.7 Ω for the first trilayer and 40±4 Ω for the second trilayer.

On all studied wafers, we have observed a factor of two larger resistance variation on the chip (die) for the 600-μA/μm$^2$ high-$J_c$ junctions than for our standard 100-μA/μm$^2$ junctions, independently of whether the high-$J_c$ junctions are on the first or the second trilayer. Since the process of junction area definition is identical for both trilayers, this can only be explained by a factor of ~2x higher barrier

TABLE I
PSE2 Fabrication Process Parameters

| Parameter / Trilayer | Nb/Al/$AlO_x$/Nb | Nb/Ni/Nb |
|---|---|---|
| Base electrode (BE) thickness (nm) | 150 | 150 |
| Top electrode thickness (nm) | 170 | 170 |
| Al thickness (nm) | 8 | - |
| Ni barrier thickness (nm) | - | 2, 2.5, 3 |
| Junction anodization voltage (V) | 20 | 20 |
| Thickness of Nb ground planes and wires (nm) | 200 | 200 |
| $SiO_2$ interlayer dielectric thickness (nm) | 200 | 200 |
| Minimum JJ diameter (nm) | 600 | 500 |
| JJ surround by BE (nm) | 350 | 350 |
| Minimum via to JJ (nm) | 350 | 350 |
| Minimum linewidth / space (μm) | 0.35 / 0.35 | 035 / 0.35 |
| Resistor layer sheet resistance, $R_s$ (Ω/sq) | 2 or 6 | 2 or 6 |
| Minimum interlayer via diameter (nm) | 400 | 400 |
| Via type | etched | etched |
| Via stacking[a] | 2 | 2 |
| $J_c$ (mA/μm$^2$) | 0.1; 0.2; 0.6 | >3 |

[a]A number of vias that can be stacked on top of each other; no stacking means that vias on adjacent levels need to be staggered.

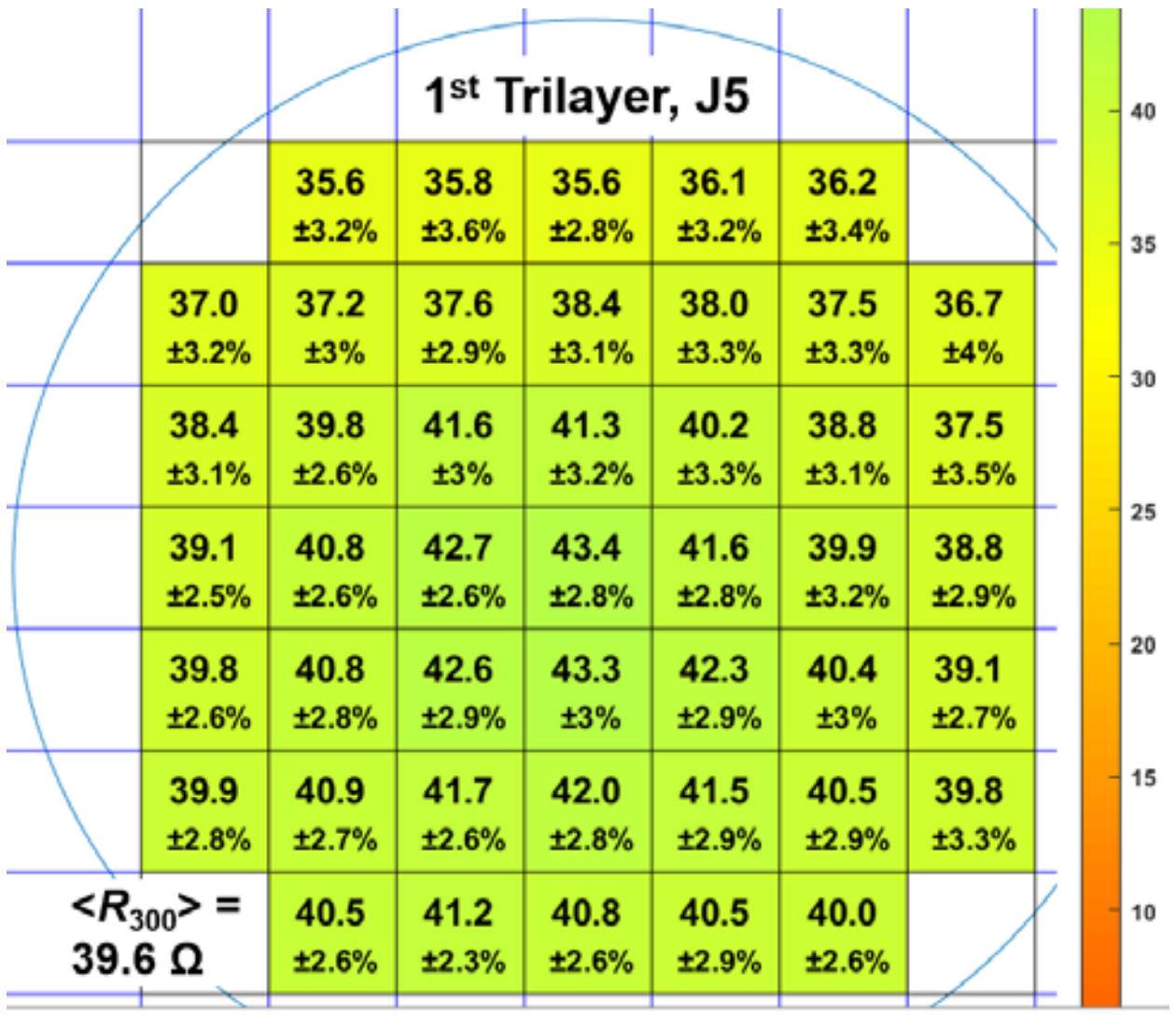

(a)

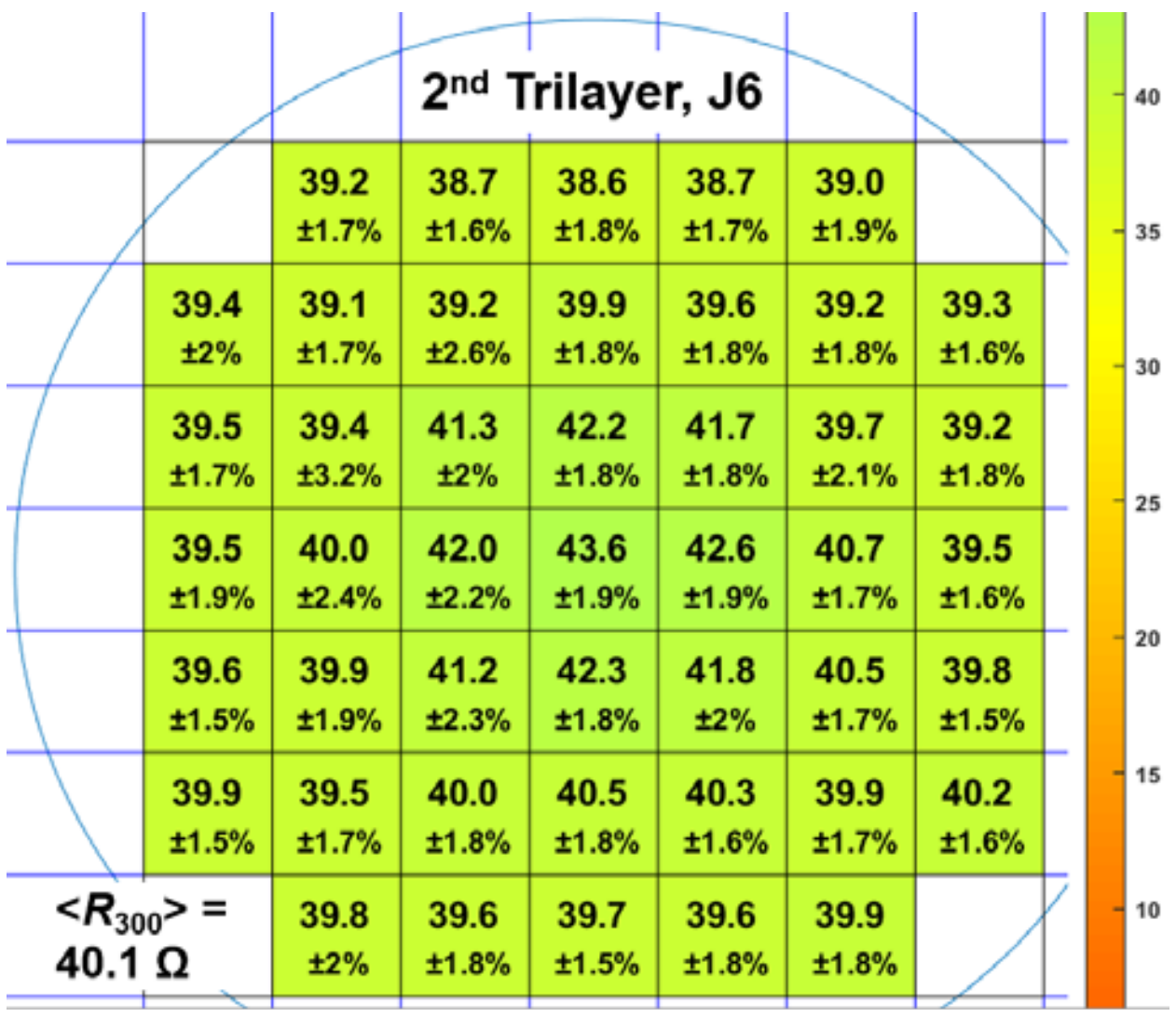

(b)

Fig. 3. Wafermaps of the mean resistance of 700-nm circular JJs at room temperature on two Nb/Al/$AlO_x$/Nb trilayers of a typical 200-mm wafer fabricated by the PSE2* process with target $J_c$ = 100 μA/μm². (a) The first junction layer J5; (b) the second junction layer J6. The target room-temperature resistance of 700-nm JJs on both trilayers is 40±4 Ω, which corresponds to $J_c$ of 100 μA/μm² ± 10%. The mean resistance on each die is shown along with the second moment of the distribution (standard deviation) normalized to the mean value. The wafer-mean values are $\langle R_{300}\rangle$ =39.6 and 40.1 Ω for J5 and J6 junctions, respectively. The total min-to-max spread across the wafer (normalized to the wafer mean) is 19.7% and 12.5%, respectively, for the first and second trilayer. The spread of $R_{300}$ is a cumulative effect of the tunnel barrier transparency, i.e., $J_c$ variation across the wafer and junction area variations.

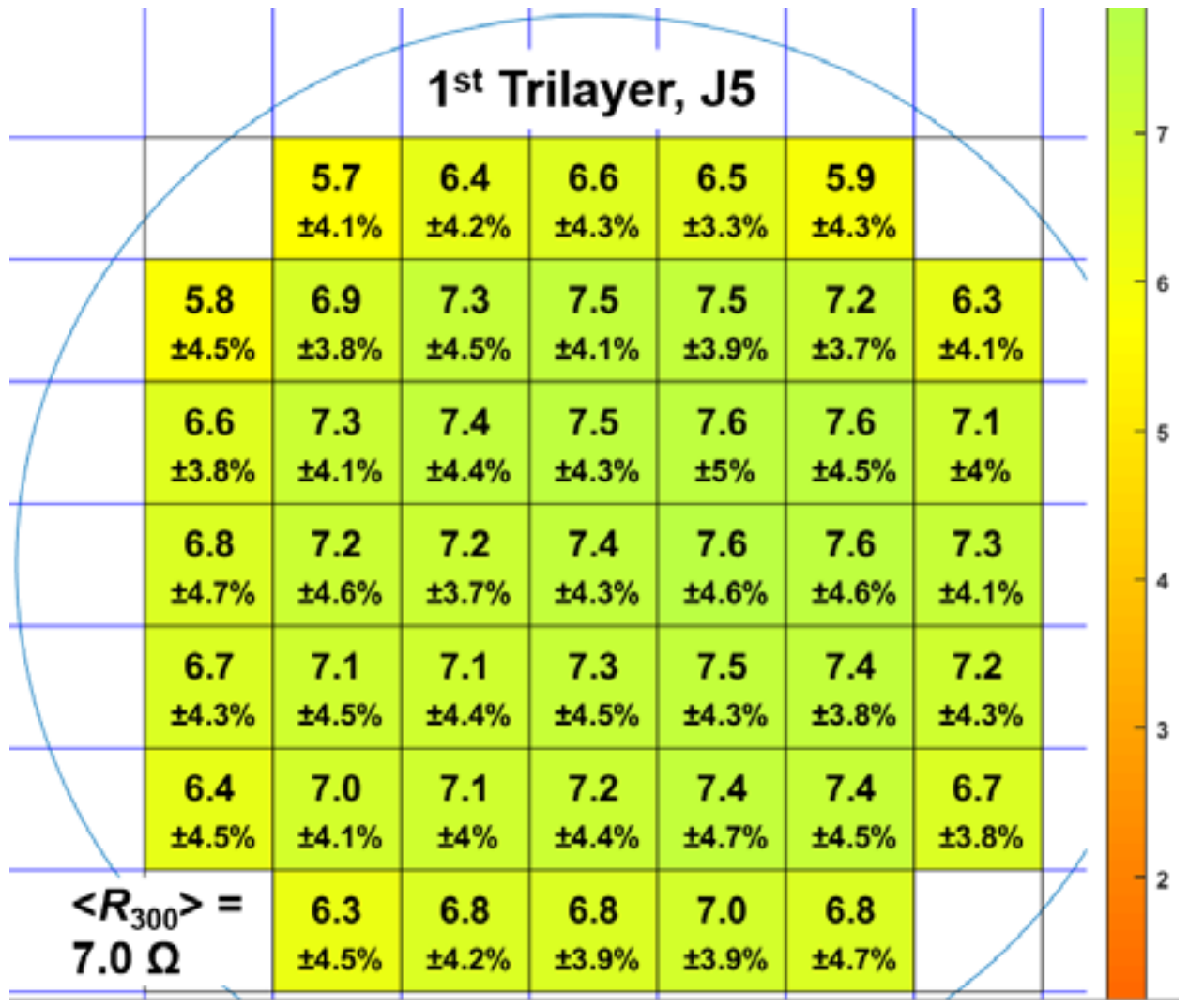

(a)

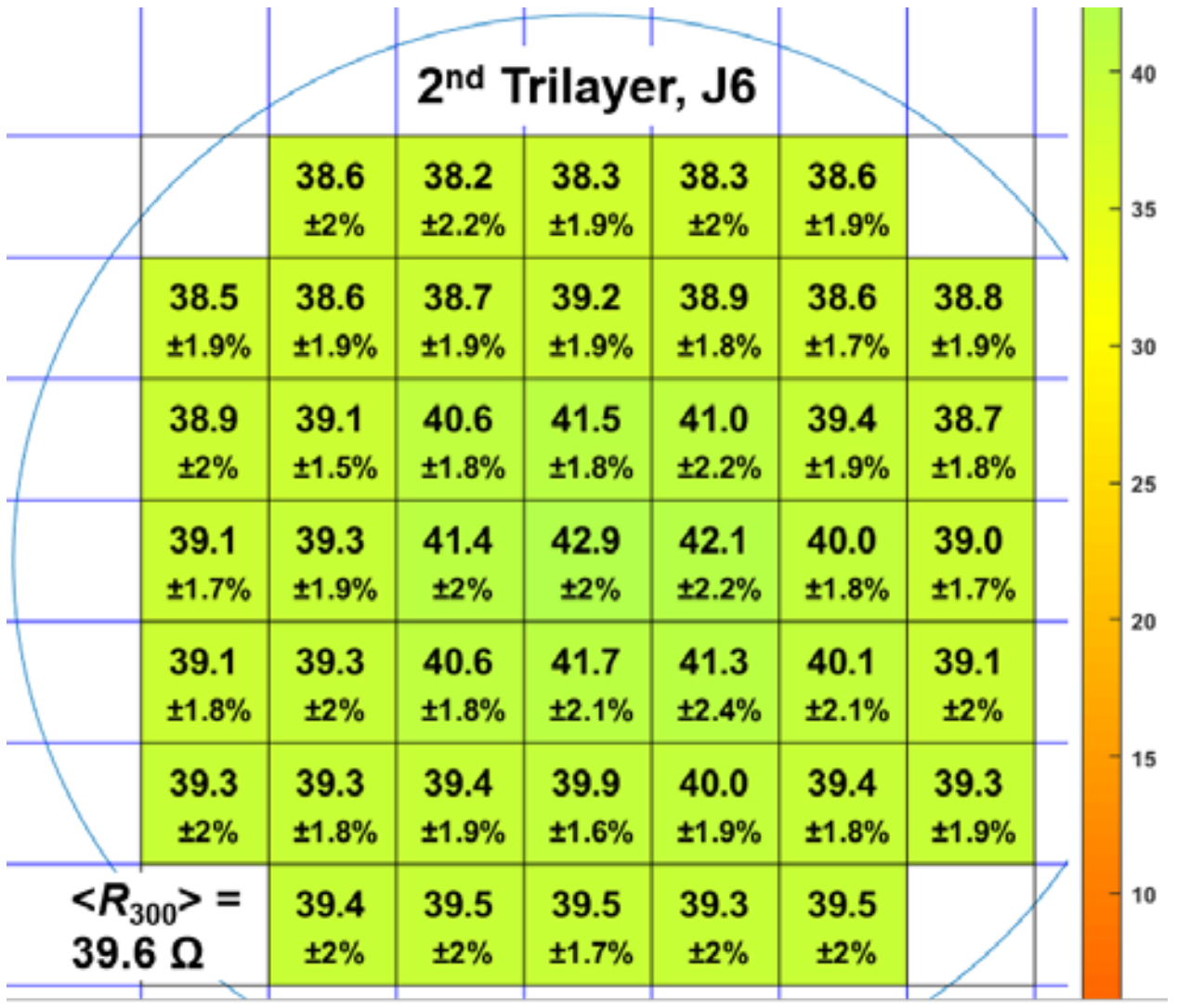

(b)

Fig. 4. Wafermaps of the mean resistance of 700-nm circular JJs at room-$T$ on two Nb/Al/$AlO_x$/Nb trilayers fabricated by the PSE2* process with target $J_c$ = 600 μA/μm² on the first trilayer and 100 μA/μm² on the second trilayer. The corresponding targets for the room-$T$ resistance of 700-nm JJs are 7±0.7 Ω on the first trilayer and 40±4 Ω on the second trilayer. The obtained wafer-mean junction resistance $\langle R_{300}\rangle$ on the trilayers is 7.0 Ω and 39.6 Ω, respectively. Note a factor of two larger total min-to-max spread of the JJ parameters across the wafer and on the die in the high-$J_c$ case than on the standard 100-μA/μm² trilayer.

transparency fluctuations in the high-$J_c$ trilayers than in the medium-$J_c$ (100-μA/μm²) trilayers.

In the truncated process PSE2*, junctions J5 and J6 share one Nb layer because the bottom electrode of the second trilayer (with J6 junctions) is used to contact the top electrode of J5 junctions on the first trilayer. For closely spaced high-$J_c$ junctions J5 and J6, switching of a JJ from the superconducting state to the normal state above the gap voltage on one JJ layer may reduce the critical current of the adjacent junction on another JJ layer. This influence is likely caused by the injection of nonequilibrium quasiparticles and phonons generated in the switched JJ into the superconducting electrode shared by the adjacent junctions, increasing the effective local temperature. These nonequilibrium effects and junction separation dependence will be considered elsewhere.

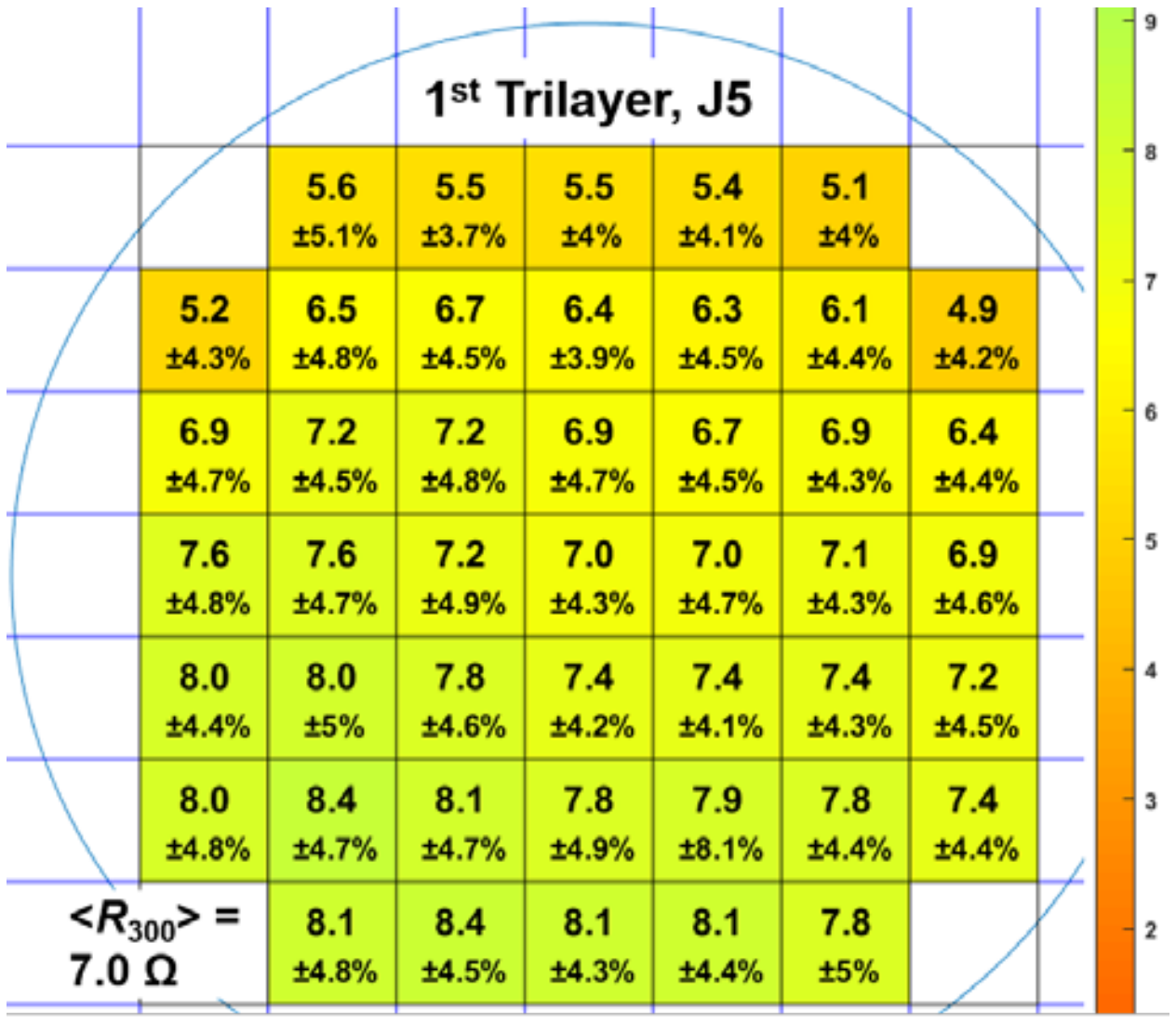

(a)

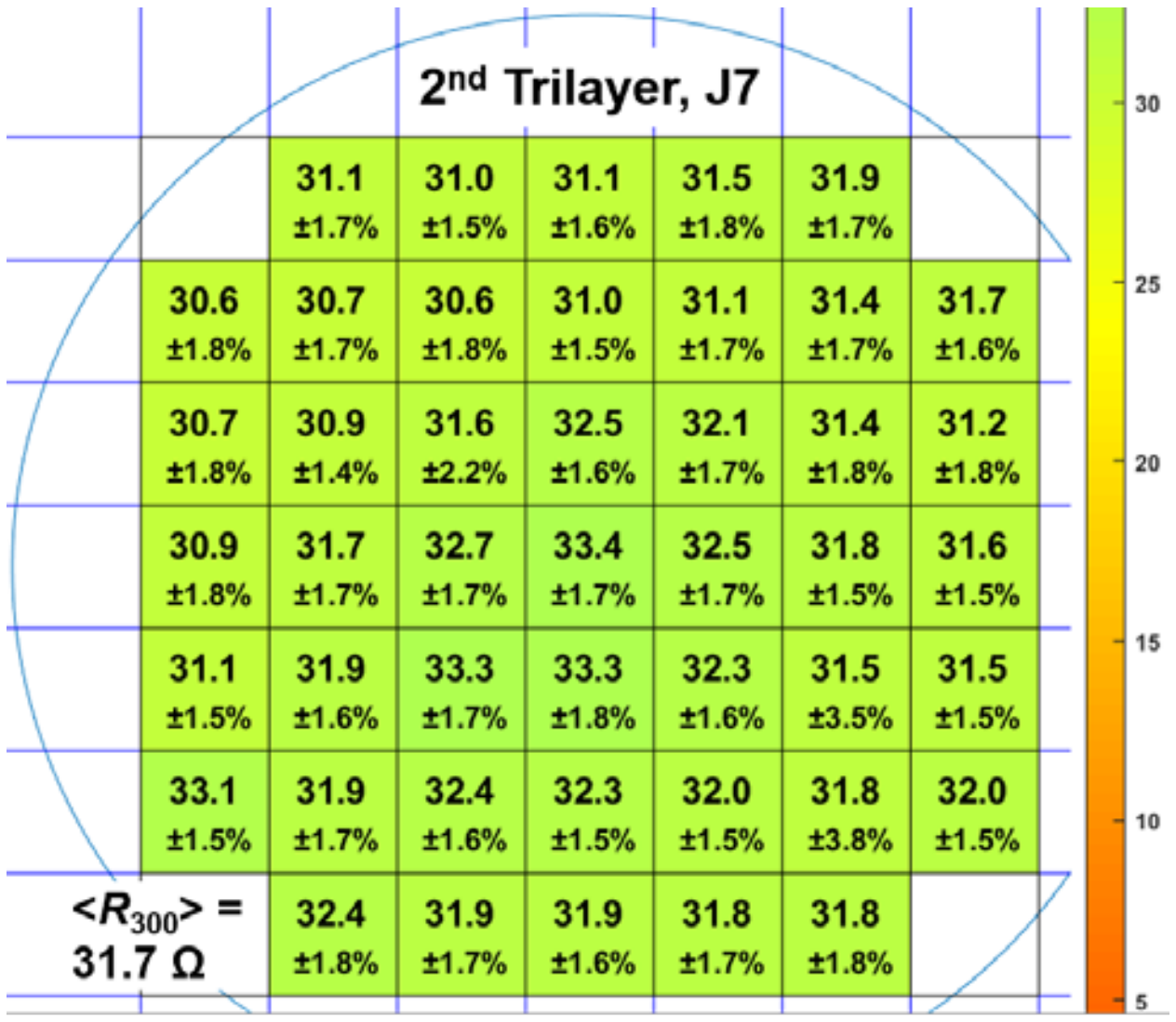

(b)

Fig. 5. Wafermaps of room-$T$ resistance of 700-nm JJs on a typical wafer fabricated in the PSE2 process. (a) The first trilayer with target $J_c$ = 600 μA/μm$^2$; (b) The second trilayer with target $J_c$ = 100 μA/μm$^2$; total min-to-max resistance variation on the second trilayer is 8.8%. The mean resistance and the standard deviation are shown on each die, based on 132 JJs measured per 5 mm x 5 mm chip. Note that the $J_c$ target on the second trilayer was missed on this wafer, resulting in the wafer-mean resistance of 31.7 Ω instead of 40 Ω.

An extra layer of Nb was added between the two layers of JJs in the full PSE2 process, as shown in Fig. 1b. Wafermaps in Fig. 5 show the mean resistance of 700-nm JJs on each die (132 JJs per 5 mm x 5 mm chip on each die) of a wafer with $J_c$ target of 600 μA/μm$^2$ on the first trilayer and the standard $J_c$ of 100 μA/μm$^2$ on the second trilayer. These results should be compared with Fig. 4. The wafer-mean resistance of 7.04 Ω and the on-chip parameter spread on the high-$J_c$ trilayer are very similar to the results in Fig. 4 for the PSE2* process. However, the min-to-max resistance variation across the wafer, and hence the $J_c$ variation on the first trilayer is larger in the PSE2 process. We continue investigating potential causes of this variation.

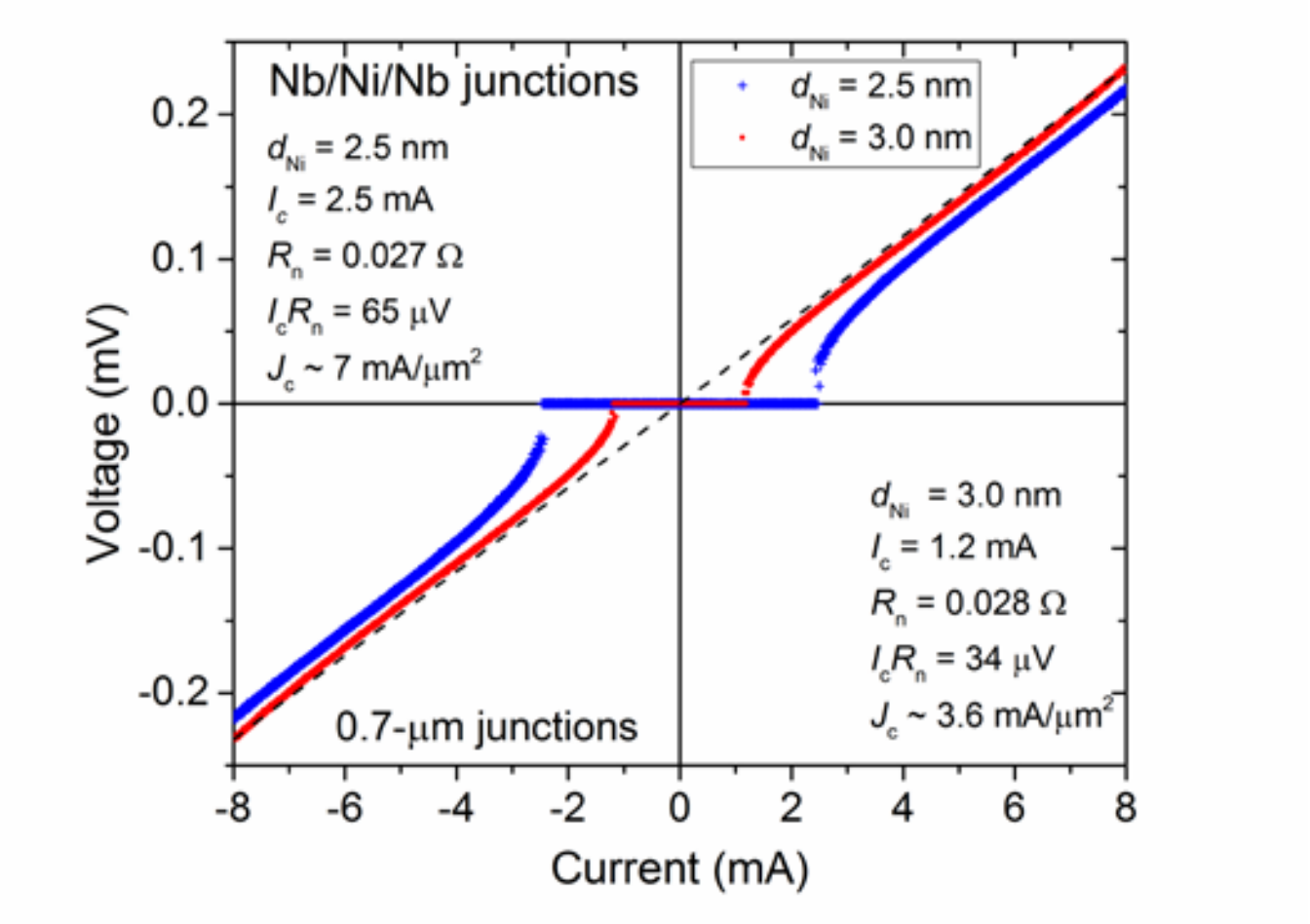

(a)

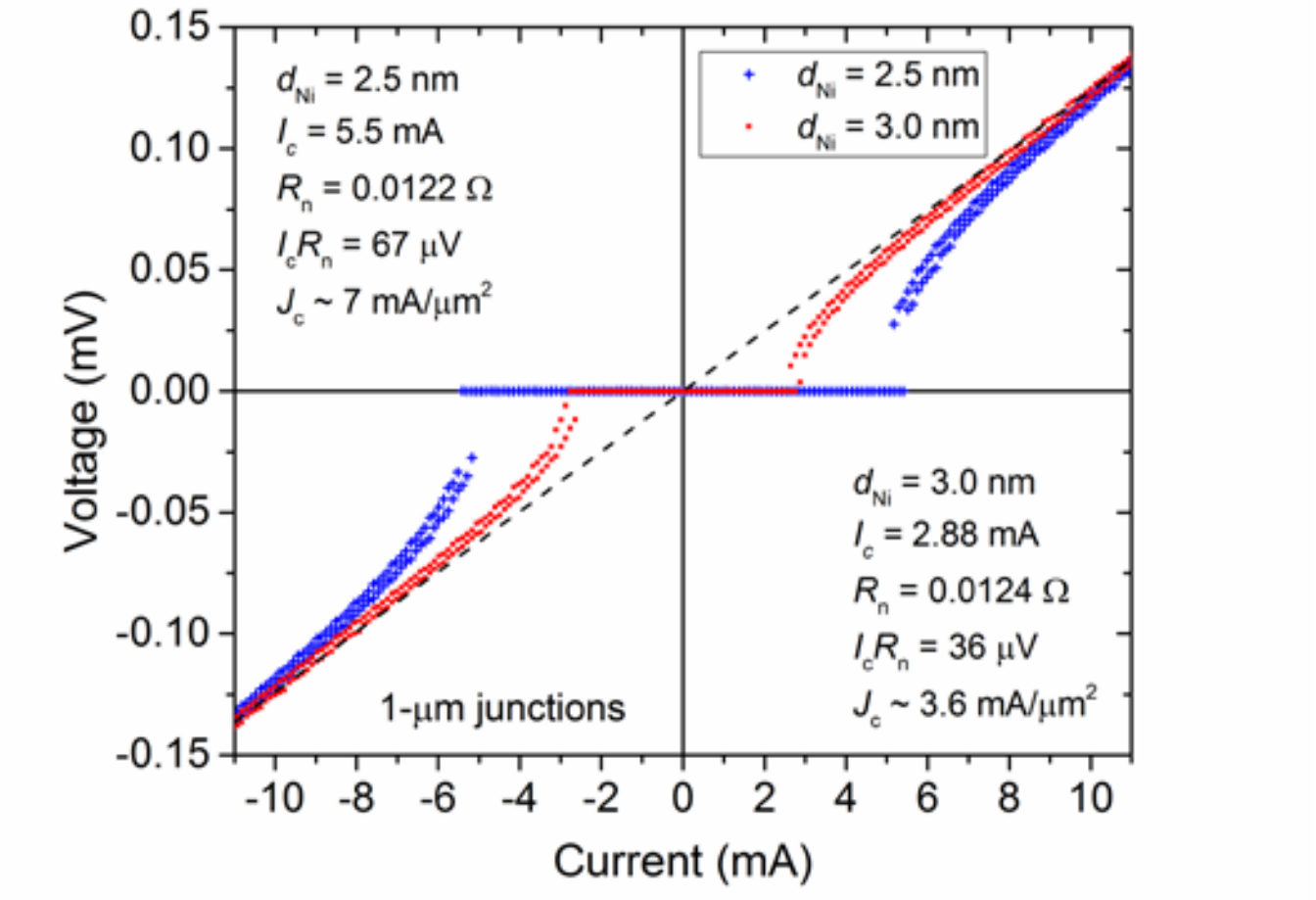

(b)

Fig. 6. Typical $I$-$V$ characteristics for Nb/Ni/Nb trilayer junctions with $t_{Ni}$ = 2.5 nm (blue crosses) and 3.0 nm (red dots) at $T$ = 4.2 K: (a) circular JJs with 0.7-μm diameter; (b) circular JJs with 1-μm diameter. All other parameters of JJs inferred from $I$-$V$ characteristics are given in the legends. The average $I_cR_n$ product of Nb/Ni/Nb JJs is about 66 μV and 35 μV for $t_{Ni}$ = 2.5 nm and 3.0 nm, respectively, which agree with the $I_cR_n(t_{Ni})$ data in [25],[26].

### B. *Nb/Ni/Nb Junctions*

The typical current-voltage ($I$-$V$) characteristics of circular Nb/Ni/Nb junctions with 0.7-μm and 1-μm diameter are shown in Fig. 6 for the trilayers with $t_{Ni}$ = 2.5 nm and 3.0 nm. JJ parameters extracted from the $I$-$V$s are also shown. For these $t_{Ni}$ values, the $J_c$ of Nb/Ni/Nb JJs exceeds 3 mA/μm$^2$. Therefore, their implementation as nonswitching π-phase shifters should not be a problem.

To demonstrate that Ni layer in the junctions is magnetic, we measured the critical current dependence on magnetic field, $I_c(B)$. For these measurements, we selected the SFS'S-type junctions because they have lower $J_c$ than Nb/Ni/Nb junctions and, respectively, lower $I_c$. This reduces self-field effects and allows using larger-diameter JJs requiring smaller magnet-ic fields to modulate $I_c$, thereby helping to reduce flux trapping.

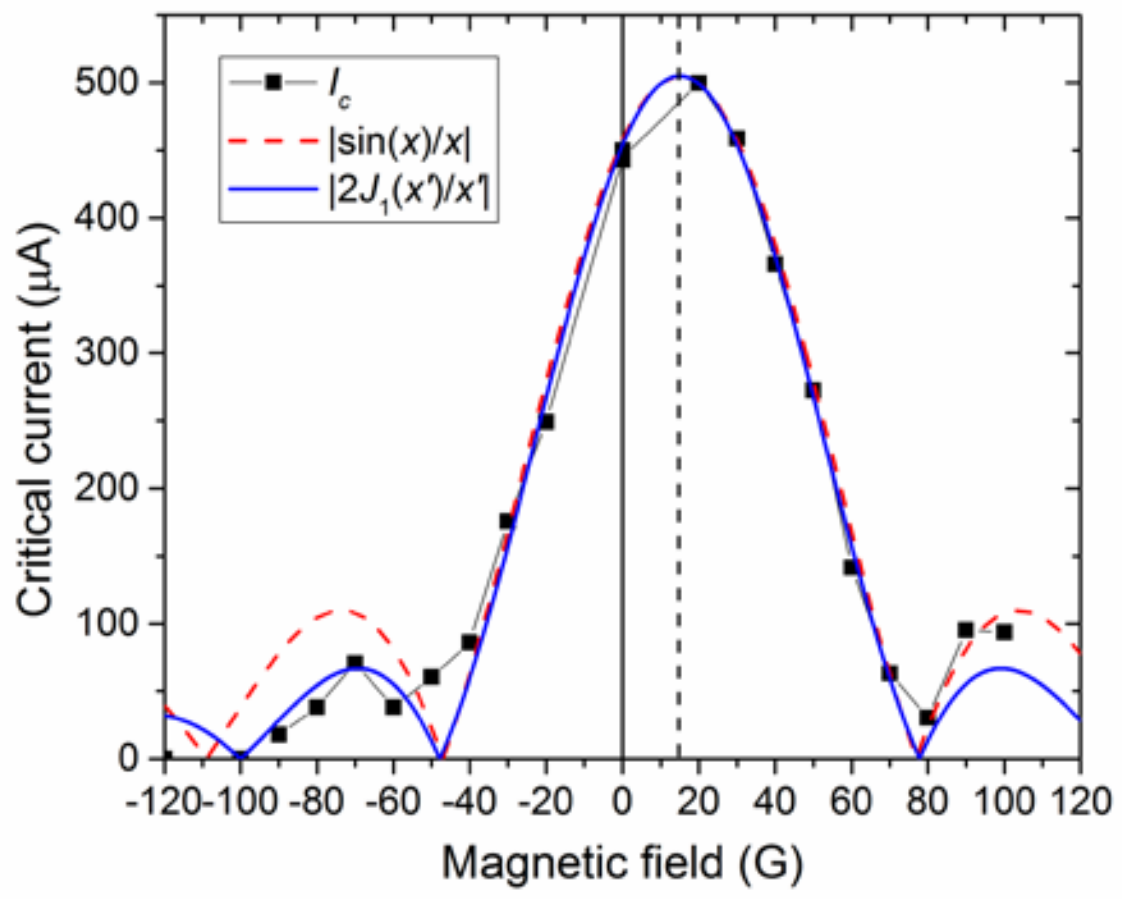

Fig. 7. Modulation of the critical current of a Nb/Ni/$Mo_2N$/Nb circular junction with $d$ = 2.24 μm and $t_{Ni}$ = 3.0 nm in an external magnetic field. The observed offset of the modulation curve is caused by magnetization of the Ni layer. The fit to the Airy diffraction pattern expected for circular junctions is shown by the blue solid curve and the fit to the Fraunhofer pattern for square junctions is the red dashed curve; see text.

Fig. 7 shows a modulation of the critical current of the typical Nb/Ni/$Mo_2N$/Nb JJ with diameter of $d$ = 2.24 μm and $t_{Ni}$ = 3.0 nm in a magnetic field $B$ parallel to the junction plane. The junction was magnetized before the measurements in a magnetic field of −100 G (0.01 T). The obtained $I_c(B)$ modulation curve is shifted with respect to the zero field by $\Delta B \approx 15$ G (1.5 mT), see Fig. 7, which corresponds to the residual internal field produced by the magnetized Ni layer.

For the circular junctions, the $I_c(B)$ dependence, in the absence of the internal magnetization, is given by the Airy diffraction pattern $I_c(B) = I_{c0}|2J_1(x)/x|$, where $J_1(x)$ is the Bessel function of the first kind and $x = \pi\Phi/\Phi_0$, $\Phi_0$ is the flux quantum, and $\Phi = B(\lambda_1 + \lambda_2 + t_{Ni})d$ is the magnetic flux threading the junction; $\lambda_1$ and $\lambda_2$ are the magnetic field penetration depths in the junction electrodes. A fit to this function is shown by the solid blue curve in Fig. 7. For comparison, the red dashed curve in Fig. 7 shows the Fraunhofer diffraction pattern, $I_c(B) = I_{c0}|\sin(x)/x|$, expected for square junctions.

From the Airy function fit, the distance between zeros in the central lobe of the $I_c(B)$ is 125.6 G, which gives $\lambda_1 + \lambda_2 \approx$ 177 nm. The latter is extremely close to the $2\lambda_{Nb}$ value of 180 nm expected for Nb electrodes, based on $\lambda_{Nb}$ = 90 nm following from the inductance measurements of Nb films used in our process [4],[28].

The residual magnetic field (remanence), $B_r$ of the Ni barrier can be estimated from $\Delta B$ by balancing magnetic fluxes induced by the external and the residual fields $\Delta B(\lambda_1 + \lambda_2 + t_{Ni}) = B_r t_{Ni}$. This gives $B_r \approx 0.9$ kG (0.09 T) and the residual magnetization of the Ni layer $M_r = B_r/4\pi$ of about 71.5 emu/cm$^3$ (8.0 emu/g or 8.0 J T$^{-1}$ kg$^{-1}$) about 1/7 of the saturation magnetization value of bulk Ni, $M_s$ = 522.7 emu/cm$^3$ (58.6 J T$^{-1}$ kg$^{-1}$) [29]. Since very few JJs have been measured so far in this way, we do not have data on the dependence of the offset field $\Delta B$ and the residual field $B_r$ on the JJ size, magnetizing field strength, and Ni layer thickness.

Unfortunately, we could not obtain sufficiently reliable information on the uniformity of the Ni barrier resistance and critical current density of Nb/Ni/Nb and Nb/Ni/$Mo_2N$/Nb trilayers on the wafer scale from room-$T$ measurements of JJs on the wafer prober. Due to a very small thickness and low resistance of the Ni barrier, the junction resistance is dominated by the parasitic resistance of Nb electrodes surrounding the junction and of the top via to the junction; see, e.g., [30] describing these parasitics. Therefore, extensive cryogenic measurements are required to assess the uniformity of the SFS and SS'FS junctions, which are planned in the future.

We also designed and fabricated various SQUID-type structures containing a combination of the regular and π-junctions on different junction layers. Modulation curves of these SQUIDs are used to demonstrate the π-phase shift introduced by each Nb/Ni/Nb JJ. These results will be presented elsewhere.

## IV. Conclusion

We have developed a new, fully planarized fabrication process PSE2 with two active layers of Josephson junctions on 200-mm wafers. Both of these layers can be Nb/Al/$AlO_x$/Nb JJs with the same $J_c$ =100 μA/μm$^2$ or the bottom layer can have a higher $J_c$, e.g., 200 or 600 μA/μm$^2$. Also, one of the two layers can be a Nb/Ni/Nb trilayer that provides high-$J_c$ Josephson π-junctions. The latter can be used as passive π-phase shifters in logic cells for increasing the integration scale of superconductor digital electronics. We presented the results of electrical characterization of both types of the junctions.

## Acknowledgment

We are very grateful to Corey Stull for his part in optimization of nickel deposition process. We are thankful to Eric Dauler, Mark Gouker, Scott Holmes, and Marc Manheimer for their interest in and support of this work.

The work was supported in part by the IARPA C3 Program. The views and conclusions contained in this publication are those of the authors and should not be interpreted as necessarily representing the official policies or endorsements, either expressed or implied, of the ODNI, IARPA, or the U.S. Government. The U.S. Government is authorized to reproduce and distribute reprints for Governmental purposes notwithstanding any copyright annotation thereon.